\documentclass{article}

\usepackage{graphics}
\usepackage{emulateapj}

\newbox\grsign \setbox\grsign=\hbox{$>$} \newdimen\grdimen \grdimen=\ht\grsign
\newbox\simlessbox \newbox\simgreatbox
\setbox\simgreatbox=\hbox{\raise.5ex\hbox{$>$}\llap
     {\lower.5ex\hbox{$\sim$}}}\ht1=\grdimen\dp1=0pt
\setbox\simlessbox=\hbox{\raise.5ex\hbox{$<$}\llap
     {\lower.5ex\hbox{$\sim$}}}\ht2=\grdimen\dp2=0pt
\def\simgreat{\mathrel{\copy\simgreatbox}}
\def\simless{\mathrel{\copy\simlessbox}}

\newcommand{\ROSAT}{\emph{ROSAT}}
\newcommand{\etal}{{et al.}\ }
\newcommand{\mr}[1]{\mathrm{#1}}

\newcommand{\ntot}{260}

\newcommand{\m}{$^{-1}$}

\newcommand{\degr}{$^\circ$}
\newcommand{\hhh}{h_{100}}
\newcommand{\mc}[1]{\multicolumn{2}{c}{{#1}}}

\newcommand{\blskip}{}

\begin{document}
\blskip

\submitted{December 6, 1999; To appear in \emph{The Astrophysical Journal}}

\title{The RASSCALS: An X-ray and Optical Study of 260 Galaxy
Groups}

\author{Andisheh Mahdavi\altaffilmark{1}} \affil{Harvard-Smithsonian
Center for Astrophysics, MS 10, 60 Garden St., Cambridge, MA 02138,
USA}

\author{Hans B\"ohringer\altaffilmark{2}} \affil{Max-Planck-Institut
f\"ur Extraterrestrische Physik, Postfach 1603, D-85740 Garching,
Germany}

\author{Margaret J. Geller\altaffilmark{3}} \affil{Harvard-Smithsonian
Center for Astrophysics, MS 19, 60 Garden St., Cambridge, MA 02138,
USA}

\and

\author{Massimo Ramella\altaffilmark{4}} \affil{Osservatorio Astronomico di Trieste, via Tiepolo 11, I-34131 Trieste, Italy}

\altaffiltext{1}{amahdavi@cfa.harvard.edu}
\altaffiltext{2}{hxb@rosat.mpe-garching.mpg.de}
\altaffiltext{3}{mgeller@cfa.harvard.edu}
\altaffiltext{4}{ramella@oat.ts.astro.it}

\lefthead{The RASSCALS}
\righthead{Mahdavi \etal}

\begin{abstract}
\blskip

We describe the \ROSAT\ All-Sky Survey---Center for Astrophysics Loose
Systems (RASSCALS), the largest X-ray and optical survey of low mass
galaxy groups to date. We draw \ntot\ groups from the combined Center
for Astrophysics and Southern Sky Redshift Surveys, covering one
quarter of the sky to a limiting Zwicky magnitude of $m_z = 15.5$. We
detect 61 groups (23\%) as extended X-ray sources.

The statistical completeness of the sample allows us to make the first
measurement of the X-ray selection function of groups, along with a
clean determination of their fundamental scaling laws. We find robust
evidence of similarity breaking in the relationship between the X-ray
luminosity and velocity dispersion. Groups with $\sigma_p < 340$ km
s\m\ are overluminous by several orders of magnitude compared to the
familiar $L_X \propto \sigma^4$ law for higher velocity dispersion
systems. An understanding of this break depends on the detailed
structure of groups with small velocity dispersions $\sigma_p < 150$
km~s\m.

After accounting for selection effects, we conclude that only 40\% of
the optical groups are extended X-ray sources. The remaining 60\% are
either accidental superpositions, or systems devoid of X-ray emitting
gas. Combining our results with group statistics from N-body
simulations, we find that the fraction of real, bound systems in our
objectively selected optical catalog is between 40\%--80\%.

The X-ray detections have a median membership of 9 galaxies, a median
recession velocity $cz = 7250$ km~s\m, a median projected velocity
dispersion $\sigma_p = 400$ km~s\m, and a median X-ray luminosity $L_X
= 3 \times 10^{42} \hhh^{-2}$ erg~s\m, where the Hubble constant is
$H_0 = 100 \hhh$ km~s\m\ Mpc\m. We include a catalog of these
properties, or the appropriate upper limits, for all 260 groups.

\end{abstract}
 
\section{Introduction}

The \ROSAT\ and \emph{ASCA} missions have shown that even low mass
systems of galaxies contain a hot intergalactic plasma. Many of the
Hickson (1982) compact groups (HCGs) are embedded in diffuse X-ray
emission detectable by the \ROSAT\ (Ebeling, Voges, \& B\"ohringer
1994; Pildis, Bregman, \& Evrard 1995; Ponman \etal 1996). Other
\ROSAT\ studies of smaller group samples (e.g. Henry \etal 1995; Burns
\etal 1996; Mulchaey \etal 1996; Mahdavi \etal 1997; Mulchaey \&
Zabludoff 1998) confirm the existence of an intergalactic plasma with
an average temperature $k T \approx 1$ keV in many loose groups.

Although heterogeneous studies abound, an objective survey of the
nearby universe for X-ray emitting groups is lacking. A search based
on a large optical catalog, drawn objectively from a three dimensional
map of the large scale structure, is essential for understanding the
physical properties of galaxy groups. Here we construct the first such
catalog. Our goals are (1) to investigate similarity breaking in the
fundamental scaling laws of systems of galaxies, (2) to make the first
calculation of the X-ray selection function of galaxy groups, and (3)
to place firm limits on the fraction of optically selected groups that
are bound.

One important example of similarity breaking is the relationship
between the X-ray luminosity $L_X$ and the average plasma temperature
$T$. Ponman \etal (1996) show that the $L_X - T$ relation is quite
steep for compact groups, with $L_X \propto T^5$, whereas for rich
clusters $L_X \propto T^3$. This result is consistent with a
``preheating'' scenario where winds from supernovae in galaxies
undergoing the starburst phase leave their mark on the poorest
systems. Such winds would deplete the intragroup plasma (Davis,
Mulchaey, \& Mushotzky 1999; Hwang \etal 1999), raise the gas entropy
relative to the gravitational collapse value (Ponman, Cannon, \&
Navarro 1999), and preferentially dim the systems with the lowest
temperatures (Cavaliere, Menci, \& Tozzi 1997).

Systems in hydrostatic equilibrium should have $T \propto \sigma_p^2$,
where $\sigma_p$ is the velocity dispersion of the dark matter halo in
which the galaxies are embedded. Thus one might expect that the $L_X -
\sigma_p$ and the $L_X - T$ relations for groups of galaxies steepen
in a similar manner.  Here we show that quite the opposite is
true. The groups with the smallest velocity dispersions are in fact
overluminous compared to the $L_X \propto \sigma^4$ law valid for
higher velocity dispersion systems. Thus the similarity breaking in
the $L_X - T$ law is apparently incommensurate with the break in the
$L_X - \sigma_p$ relation. We discuss several plausible explanations
for this lack of concordance.

The paper is organized as follows. After constructing the catalog (\S
2), we examine the $L_X - \sigma_p$ relation (\S 3), calculate the
selection function (\S 4), discuss the $L_X - \sigma_p$ flattening (\S
5), and summarize our findings (\S 6). We call our groups the \ROSAT\
All-Sky Survey---Center for Astrophysics Loose Systems, or RASSCALS.

\section{Data}

\subsection{Optical Group Selection}

We extract the optical group catalog for the RASSCALS study from two
complete redshift surveys. Our catalog includes a wide variety of
systems, from groups with only $5$ members to the Coma cluster
\footnote[5]{In a previous work (Mahdavi \etal 1999), we referred to
the Center for Astrophysics--SSRS2 Optical Catalog (CSOC) as a
distinct entity from the RASSCALS, which was to be the X-ray
catalog. We no longer make that distinction, and refer to the
X-ray/optical catalog simply as the RASSCALS.}.

The Center For Astrophysics Redshift Survey (Geller \& Huchra 1989;
Huchra \etal 1990; Huchra, Geller, \& Corwin 1995; CfA) and the
Southern Sky Redshift Survey (Da Costa \etal 1994; Da Costa \etal
1998), both complete to a limiting Zwicky magnitude $m_z \approx
15.5$, serve as sources for the RASSCALS. The portion of the surveys
we use covers one fourth of the sky in separate sections described in
Table \ref{tbl:skyportions}. We transform the redshifts to the Local
Group frame ($\Delta c z = 300 \sin{l} \cos{b}$), and correct them for
infall toward the center of the Virgo cluster (300 km~s\m\ towards
$\alpha_{2000} = 12^\mr{h} 31.2\mr{m}$, $\delta_{2000} = 12$\degr
2.54$^\prime$).

We use the two-parameter friends-of-friends algorithm (FOFA) to
construct the optical catalog. Huchra \& Geller (1982) first described
the FOFA for use with redshift surveys, and Ramella, Pisani, \& Geller
(1997) applied it to the NRG data. The FOFA is a three-dimensional
algorithm which identifies regions with a galaxy overdensity $\delta
\rho / \rho$ greater than some specified threshold. A second fiducial
parameter, $V_0$, rejects galaxies in the overdense region which are
too far removed in velocity space from their nearest neighbor. The
N-body simulations of Frederic (1995) and Diaferio (1999) show that
the Huchra \& Geller (1982) detection method misses few real systems,
at the cost of including some spurious ones. We apply the FOFA to the
combined NRG, SRG, and SS2 redshift surveys with $\delta \rho/\rho =
80.$
 
The RASSCALS optical catalog contains \ntot\ systems with $n\ge 5$
members and 3000 km~s\m $ \le c z \le 12000$ km~s\m. The low velocity
cutoff rejects systems that cover a large area on the sky and thus may
be affected by the Local Supercluster. The median recession velocity
for the systems is 7000 km~s\m; the effects of cosmology and evolution
are negligible throughout the sample. Table \ref{tbl:rasscals} lists
the individual groups and their properties.  Figures
\ref{fig:picstart}--\ref{fig:picend} show the sky positions of the
member galaxies for the systems with statistically significant
extended X-ray emission in the RASS.

To compare the membership of groups which have different redshifts we
also compute $n_{17}$, the number of group members brighter than an
absolute magnitude $M_z = -17$, corresponding to $m_z = 15.5$ for a
group at $cz = 3200$ km~s\m. To calculate $n_{17}$, we assume that the
galaxies in groups have the same luminosity function as the Center for
Astrophysics Redshift Survey (Marzke \etal 1994), reconvolved with the
magnitude errors. The resulting distribution is well-represented by a
Schechter (1976) function with a characteristic absolute magnitude
$M_* = -19.1$ and a faint-end slope $\alpha = -1$.  Table
\ref{tbl:rasscals} lists $n_{17}$, which has a median value of 44.

\subsection{X-Ray Field Selection}

For every system in the RASSCALS optical catalog, we obtain X-ray data
from a newly processed version of the \ROSAT\ All-Sky Survey (Voges
\etal 1999), which corrects effects leading to a low detection rate in
the original reduction.

We first assign each system a seven-character name, beginning with
``NRG,'' ``SRG,'' or ``SS2,'' followed by ``b'' or ``s'' (specifying
the angular size of the system as ``big,'' with $c z < 8500$ km~s\m\
or ``small,'' with $c z > 8500$ km~s\m, respectively), followed by a
three-digit number.

For each ``small'' system we extract a square field measuring $2^\circ
\times 2^\circ$ from the RASS; for the ``big'' systems we extract a
$3.5^\circ \times 3.5^\circ$ square. The fields are centered at the
mean RA and DEC of the galaxies; every field is at least large enough
to include a circle with a projected radius of $1\hhh^{-1}$ Mpc around
the optical center of the system it contains.  We use photons in the
0.5--2.0 keV hard energy band of the Position-Sensitive Proportional
Counter (PSPC channels 52-201).

\subsection{Detection Algorithm}

The X-ray detection algorithm consists of four steps: measurement of
the background, source identification, decontamination, and
measurement of the source flux.

We determine the mean background by temporarily rebinning the
exposure-weighted \ROSAT\ field into an image with $15 \arcmin$
pixels. We clean the image of all fluctuations with an iterative,
$2.5\sigma$ clipping algorithm.  The adopted background is then the
average of the remaining pixel values.

To estimate the probability that a given group is an X-ray source, we
use an optical galaxy position template (GPT). Mahdavi \etal (1997),
who search the RASS for X-ray emission from a small subset of our
sample, describe this method in greater detail. The GPT is defined as
the union of all projected $d = 0.2\hhh^{-1}$ Mpc regions around the
group members, excluding any galaxies isolated by more than $d$ from
the rest of the group. We count the X-ray photons within the GPT and
evaluate the probability that they are drawn from the background
distribution. All groups that have a detection significance greater
than $2.5\sigma$ progress to the next step.
 
We identify the emission peak which coincides most closely with the
optical center of the group as its X-ray counterpart, and calculate
the X-ray position of the group with the intensity-weighted first
moment of the pixel values.  Using standard maximum likelihood
techniques, we identify contaminating X-ray point sources over the
entire field. We remove these sources by excising a ring of radius
3$\arcmin$, roughly three times the full width at half maximum of the
\ROSAT\ PSPC point spread function (PSF). Unrelated extended sources
often contaminate the group emission; we use a suitably larger
aperture to remove them. We have examined publicly available ROSAT
High-Resolution Imager (HRI) observations of a few groups, and find
that our RASS decontamination procedure is satisfactory. 

To reject groups with entirely pointlike X-ray emission, we calculate
$N(R)$, the cumulative distribution of the ROSAT counts. We use use
the Kolmogorov-Smirnov (KS) Test to compare the shape of the emission
peak with that of the PSPC PSF combined with the background. We take
sources with $P_\mr{KS} \le 0.05$ as inconsistent with the PSF.
	
Finally, we convert the PSPC count rate into $L_X(R)$, the 0.1--2.4
keV X-ray luminosity contained within a ring of projected radius $R$.
The Appendix describes the flux conversion procedure in detail.

\vspace{0.2in}
\resizebox{3in}{!}{\includegraphics{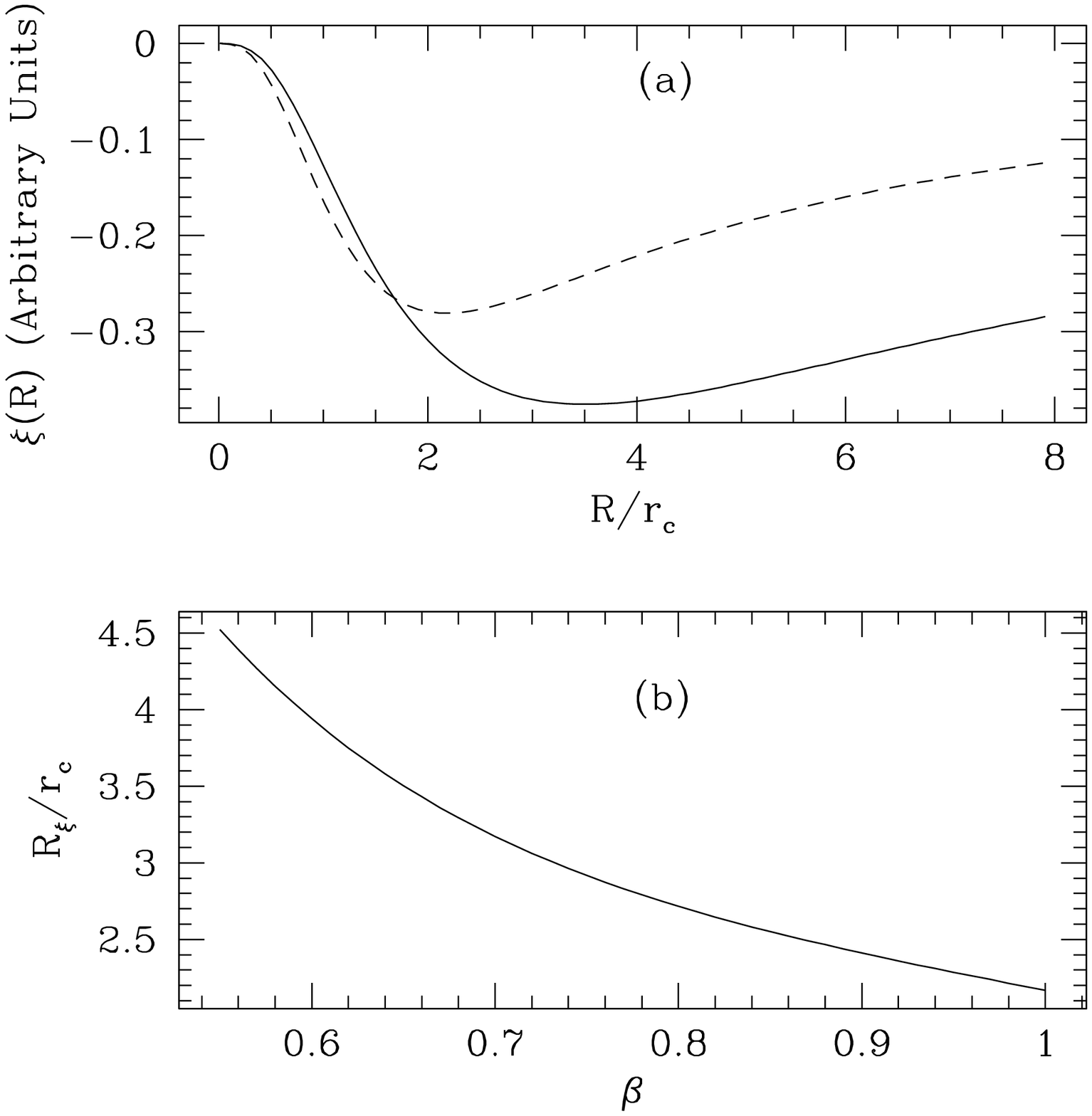}}
\figcaption{Properties of the NOCORE estimator: (a) $\xi(R)$ for
$\beta$-models with $\beta = 0.65$ (solid line) and $\beta = 1$
(dashed line); (b) $R_\xi$, the radius where $\xi(R)$ is
minimum, as a function of $\beta$. \label{fig:betacore}}
\vspace{0.2in}

\subsection{Core Radius Estimation}
\label{sec:nocore}

Here we describe a procedure for identifying a physical scale for the
X-ray emission. There is a great deal of evidence that the emissivity
profiles of clusters of galaxies exhibit a characteristic scale, or
core radius, $r_c$ (e.g. Jones \& Forman 1984; Mohr, Mathiesen, \&
Evrard 1999). For example, the ``$\beta$-model'' emissivity profile
frequently used to fit observations of dynamically relaxed systems,
\begin{equation}
\epsilon(r) \propto \left( 1 + \frac{r^2}{r_c^2} \right)^{-3 \beta},
\end{equation}
is nearly constant for physical radii $r \ll r_c$, and scales as
$r^{-6 \beta}$ for $r \gg r_c$. There is also evidence for cuspy
profiles ($\epsilon \propto r^{-1}$ for $r \ll r_c$) in clusters with
cooling flows (Thomas 1998).

The usual method for measuring $r_c$ from X-ray observations consists
of projecting $\epsilon(r)$ along one dimension, and fitting the
resulting surface brightness profile to the data. This approach has
the disadvantage that the resulting estimate of $r_c$ is
model-dependent, and is often strongly correlated with the slope
parameter $\beta$, even with very high quality data (e.g. Jones \&
Forman 1984; Neumann \& Arnaud 1999). Furthermore, its application to
the RASSCALS is limited, because the small number of counts and the
relatively large uncertainty in the background make it difficult to
reconstruct accurate surface brightness profiles for all but the
brightest systems.

We therefore use the Nonparametric Core Radius Estimator (NOCORE;
Mahdavi 2000) to avoid the core fitting procedure and its associated
uncertainties. NOCORE is model-independent; it does not require an
 estimation of the background, and it relies on the properties
of the integrated emission profile, rather than the differential
profile, to estimate the core radius. Its only assumption is the
constancy of the background level at the position of the object
of interest.

Now we outline the procedure. Consider the measured count rate within
an annulus $R$ from the X-ray center of the group: it consists of the
emission of the group itself, $S(R)$, plus the constant background
count rate per unit area, $B$:
\begin{equation}
N(R) = S(R) + \pi R^2 B.
\end{equation}
The fundamental basis of NOCORE is the observation that the quantity $
N(R) - k^2 N(R/k) $, where $k$ is a number greater than 1, is
completely independent of the constant background. Formally, we define
the NOCORE radius as the radius where the function
\begin{equation}
\xi(R)  \equiv  \frac{N(R) - 4 N(R/2)}{R}
\end{equation}
has a global minimum. The division by $R$ is necessary to obtain a
detectable minimum.

Figure \ref{fig:betacore}a shows $\xi(R)$ for theoretical
$\beta$-models with $\beta = 0.65$ and $\beta = 1$. The minimum,
$R_\xi$, is well-defined in both cases. The location of $R_\xi$ as a
function of $\beta$ appears in Figure \ref{fig:betacore}b. We have
carried out numerical tests of the method, adding Poisson noise and a
background to a variety of $\beta$-models, to verify that we recover
the appropriate core radius without bias. When applying the method to
the observations, we use bootstrap resampling to determine 68\%
confidence intervals on $R_\xi$.

As long as there is a characteristic scale in the emissivity profile
of a system, NOCORE will find it. The function $\xi(R)$ has a
well-defined minimum even in cases when the $\beta$-model is not a
good description of the emissivity, for example systems with a cooling
flow. If there is more than one characteristic scale in the
profile---if the emissivity has features at several radii, because of
substructure in the cluster, for example---then $\xi(R)$ has more than
one minimum. If the profile is a pure power law, NOCORE shows no core;
$\xi(R)$ is then a monotonically decreasing or increasing function of
$R$. 

We use the radius at which $\xi(R)$ is minimum as a measure of the
physical scale of the X-ray emission.

\begin{figure*}
\resizebox{7in}{!}{\includegraphics{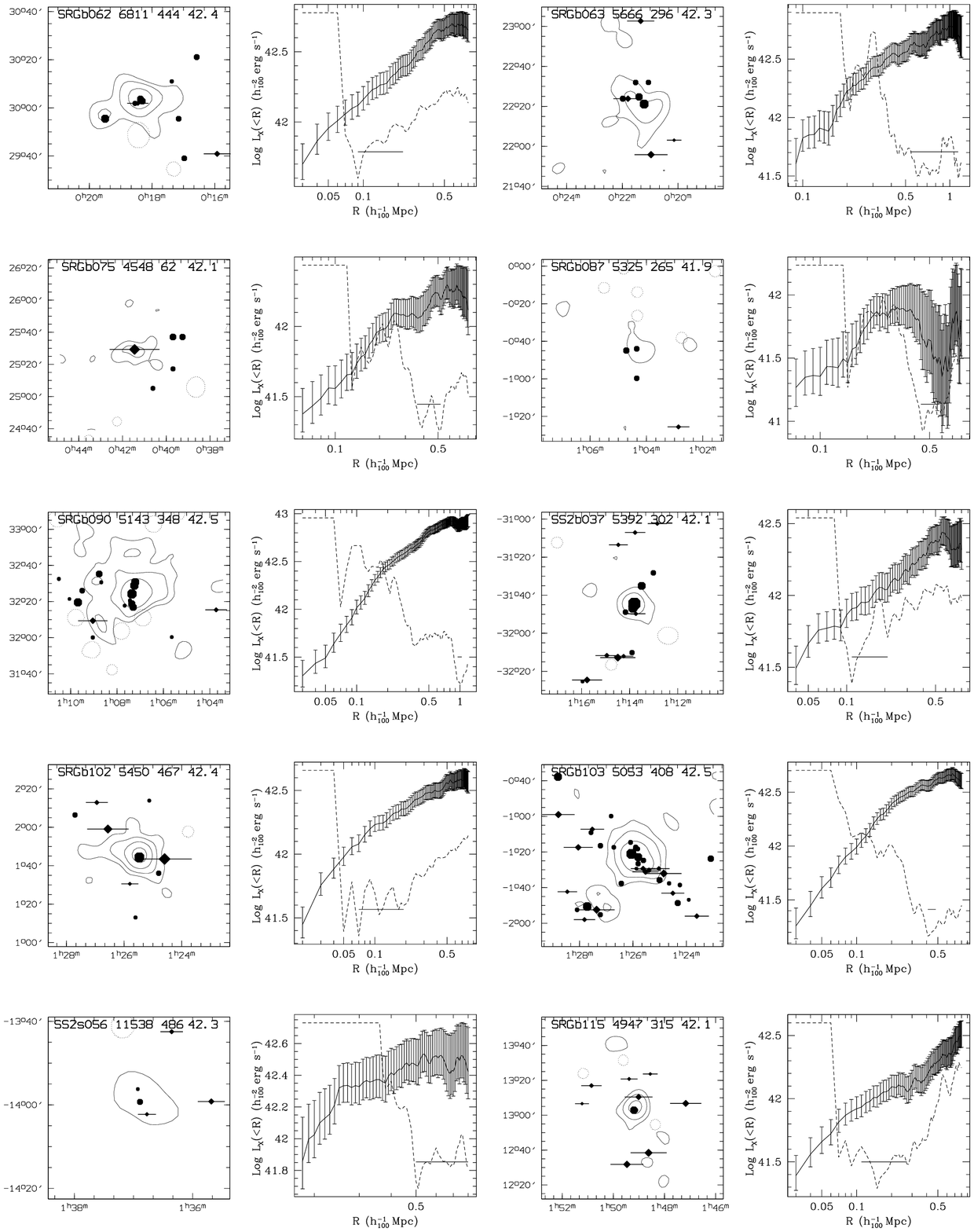}}
\figcaption[picture01.ps]{The RASSCALS. Sky maps are 1.5$\hhh^{-1}$
Mpc wide, and include the name of each group, its mean recession
velocity, velocity dispersion, and log X-Ray luminosity within
0.5$\hhh^{-1}$ Mpc. The solid X-ray emission contours begin at
1$\sigma$ above the background and increase by a factor of 2.  Dotted
regions represent sources excised by the decontamination algorithm.
Circles represent early-type galaxies, and the elongated symbols
represent late-type galaxies. The size of the symbols is proportional
to their apparent Zwicky magnitude; the largest symbol corresponds to
$m_Z = 12$, and the smallest corresponds to $m_Z = 15.5$. Also shown
are the background-subtracted cumulative X-ray luminosity profile, and
the NOCORE estimator $\xi(R)$. The 68\% confidence interval of the
NOCORE radius $R_\xi$ appears as a horizontal bar.
\label{fig:picstart}}
\end{figure*}

\begin{figure*}
\resizebox{7in}{!}{\includegraphics{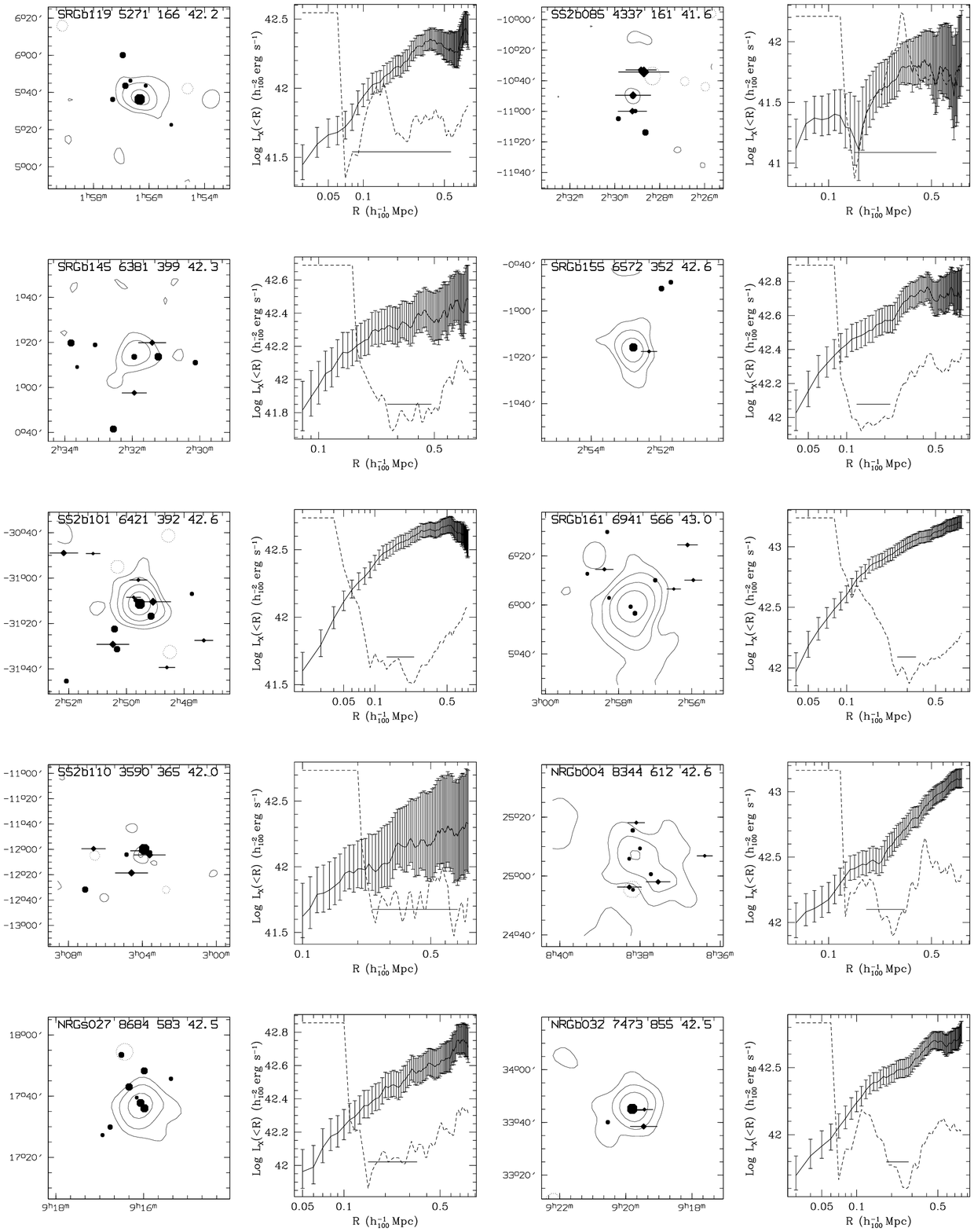}}
\figcaption[picture02.eps]{See caption for Figure
\protect\ref{fig:picstart}. Figures for the remainder of the X-ray
detected RASSCALS will appear in the published version of this
article, in \emph{The Astrophysical Journal}.
\label{fig:picend}}
\end{figure*}

\subsection{General Properties of the Final Catalog}

Table \ref{tbl:detailed} lists the projected velocity dispersion,
$\sigma_p$, the 0.1--2.4 keV X-ray luminosity, and the NOCORE radius
$R_\xi$ of the detected groups. There are 61 detections, of which two
(NRGs372 and NRGs392) are bright X-ray clusters (Abell 2147 and 2199)
that the friends-of-friends algorithm has broken up into pieces. We
count these clusters as detections but we do not calculate
luminosities for them. Figures \ref{fig:picstart}--\ref{fig:picend}
show galaxy positions and X-ray emission contours for the detected
groups.

Because the diffuse X-ray emission is generally a marker of gas held
in a gravitational potential, these systems are probably bound
configurations. There is, however, a chance that the X-ray emission
might be due to projection along an unbound filament in the large
scale structure of the universe (Hernquist, Katz, \& Weinberg
1995). The galaxies projected along this line of sight might also have
similar redshifts without being bound. However, deeper redshift
surveys in the fields of X-ray emitting RASSCALS show that a number of
these systems have velocity dispersion profiles $\sigma_p(R)$ that
decline as a function of projected distance from the group center;
these profiles are consistent with the expectation for a relaxed
dynamical system (Mahdavi \etal 1999).

In Figures \ref{fig:picstart}--\ref{fig:picend} we also show the
cumulative luminosity profile and the NOCORE estimator $\xi(R)$. In
several cases $\xi(R)$ appears to have several local minima in
addition to the global minimum. The local minima are almost always due
to Poisson noise and deviations from spherical symmetry in the
structure of the gas. The confidence intervals on $R_\xi$ take these
fluctuations into account: when the fluctuations in $\xi(R)$ dominate
its shape, the error in $R_\xi$ is large. But when $\xi(R)$ has a
well-defined global minimum and the fluctuations are small, $R_\xi$ is
relatively well determined.

\section{$L_X - \sigma_p$ Relation}
\label{sec:lxsig}

Here we examine the relationship between the X-ray luminosity and the
projected velocity dispersion. First we comment on the link between
the $L_X - \sigma_p$ scaling law, which relates X-ray and optical
data, and the $L_X - T$ scaling law, which is internal to X-ray
data. Then we describe the actual relation.

\subsection{Background on the Scaling Laws}
\label{sec:background}

If the member galaxies trace the total mass distribution in a cluster,
a simple theoretical calculation (Quintana \& Melnick 1982) predicts
that a spherically symmetric ball of gas should have $L_X \propto
f^2 \sigma_p^3 T^{1/2}$, where $f$ is the ratio of the gas mass to the
total mass. A further, common assumption is that $T \propto
\sigma_p^2$, i.e., that the emission-weighted gas temperature is
proportional to the depth of the gravitational potential. These
assumptions yield $L_X \propto f^2 T^2 \propto f^2 \sigma^4$.

The observed $L_X - \sigma_p$ relation for rich clusters is in good
agreement with the theoretical prediction; Quintana \& Melnick (1982)
and Mulchaey \& Zabludoff (1998), for example, find slopes consistent
with $L_X \propto \sigma^4$. The empirical $L_X - T$, relation,
however, is somewhat steeper than expected, with most finding $L_X
\propto T^{2.75}$, even after removing the central cooling flow region
(e.g., Markevitch 1998 and references therein).  

If the discrepancy between the simple theoretical prediction, $L_X
\propto T^2$, and the observations is real, several effects might
explain it. It could be that $f$ increases slightly with $T$ (David,
Jones, \& Forman 1995), or that $T \propto \sigma^{1.5}$, consistent
with a nonisothermal, polytropic gas distribution (Wu, Fan, \& Xu
1998). Finally, preheating of gas in the $k T \simless 4 $ keV systems
may preferentially dim them, leading to a steeper relation. Ponman,
Cannon, \& Navarro (1999) show that this latter possibility is
particularly attractive because it also accounts for differences in
the shapes of X-ray surface brightness profiles among $k T \simless 4$
keV and $k T \simgreat 4$ keV clusters. Cavaliere \etal (1997) work
out the $L_X - T$ relation for this scenario, and find that it
steepens gradually as $T$ declines, with $L_X \propto T^5$ for poor
groups , $L_X \propto T^3$ for 2 keV $\simless k T \simless$ 7 keV
systems, and $L_X \propto T^2$ for the hottest clusters. This $L_X -
T$ relation fits temperatures and luminosities for a range of systems
from poor groups to clusters.

Now, if a single power law describes the scaling of the velocity
dispersion $\sigma_p$ with the temperature $T$, and the Cavaliere
\etal (1997) preheating model is correct, one should observe a
similarly steep $L_X - \sigma_p$ relation for poor groups.  Three
different works have attempted a measurement of the faint end of this
relation, with three different results.

\begin{enumerate}
\item Ponman \etal (1996) analyze a mixture of pointed and RASS
observations of a sample of Hickson (1982) Compact Groups (HCGs
hereafter). They obtain $L_X \propto \sigma_p^{4.9 \pm 2.1}$ for the
groups with pointed observations. While this result favors a steeper
slope, the 68\% confidence interval is quite large: the HCGs contain
as few as three member galaxies, and hence it is very difficult to
estimate the correct velocity dispersion. Also, HCGs are often
embedded in much richer systems (Ramella \etal 1994), and this
embedding may further bias the value of the velocity dispersions.

\item Mulchaey \& Zabludoff (1998; MZ98 hereafter) carry out deep
optical spectroscopy for a more limited sample of poor groups with
pointed \ROSAT\ observations. Because they obtain $\approx 30$ members
per group, their derived velocity dispersions should be more reliable
than those of Ponman \etal (1996). They obtain $L_X \propto
\sigma^{4.3 \pm 0.4}$ for a combined sample of groups and clusters. 

\item Mahdavi \etal (1997) use our method to examine \ROSAT\ data for
a small but statistically complete subset of the RASSCALS. They do
not, however, excise emission from individual galaxies; furthermore,
they assume a constant plasma temperature $k T = 1$ keV, rather than
leaving $T$ free to vary as we do here. They obtain $L_X \propto
\sigma_p^{1.56 \pm 0.25}$, much shallower than either Ponman \etal
(1996) or MZ98.

\end{enumerate}

In summary, Ponman \etal (1996) find an $L_X - \sigma_p$ relation
 consistent with the simplest predictions of preheating models;
MZ98 derive a relation that is consistent with the standard picture
with no preheating; and Mahdavi \etal (1997) find that the faint-end
slope is much shallower than the prediction of either of the two
scenarios.

We now consider the $L_X - \sigma_p$ relation for the complete set
of RASSCALS. Our procedure differs from that of Mahdavi \etal (1997),
because we remove contaminating sources whenever they are detectable,
model the plasma temperature, and use an updated version of the RASS.

To compare the RASSCALS $L_X - \sigma_p$ relation with that of richer
systems, we take cluster X-ray luminosities from the paper by
Markevitch (1998), where cooling flows are removed from the
analysis. We use only clusters which have velocity dispersion listed
in Fadda \etal (1996), who consider systems with $\ge 30$ measured
redshifts. Table \ref{tbl:clusters} lists these data. Figure
\ref{fig:lxsig} shows the combined cluster-RASSCALS data. The $L_X -
\sigma_p$ seems to flatten as the luminosity decreases.

\subsection{Details of the Fitting Procedure}
\label{sec:fitting}

To place a quantitative constraint on the degree of flattening,
we fit a broken power law of the form
\begin{eqnarray}
\log{\frac{L_X}{L_k}} & = & 
    s(\sigma_p,\sigma_k)\log{\frac{\sigma_p}{\sigma_k}};  \\
s(\sigma_p,\sigma_k) & = & \left\{
\begin{array}{cl}
s_1 & \mr{if\ }\sigma_p < \sigma_k \\
s_2 & \mr{if\ }\sigma_p > \sigma_k
\end{array} \right.
\end{eqnarray}
Here $s_1$ and $s_2$ are the faint-end and bright-end slopes, 
respectively, and $(\sigma_k,L_k)$  is the position of the
knee of the power law. We then minimize a merit
function appropriate for data with error in two
coordinates (Press \etal 1995, \S 15.3),
\begin{equation}
\chi^2 = \sum_{i=1}^n \frac{\left[\log{(L_i/L_k)} - 
s(\sigma_i,\sigma_k)\log{(\sigma_i/\sigma_k)} \right]^2}
{\left(\Delta \log{L_i}\right)^2 + s(\sigma_i,\sigma_k)^2
 \left(\Delta \log{\sigma_i}\right)^2 },
\end{equation}
where $(\sigma_i,L_i)$ are the measurements, with errors $(\Delta
\sigma_i,\Delta L_i)$. To minimize the $\chi^2$ we apply the following
procedure.

\begin{figure*}
\begin{center}
\resizebox{!}{8in}{\includegraphics{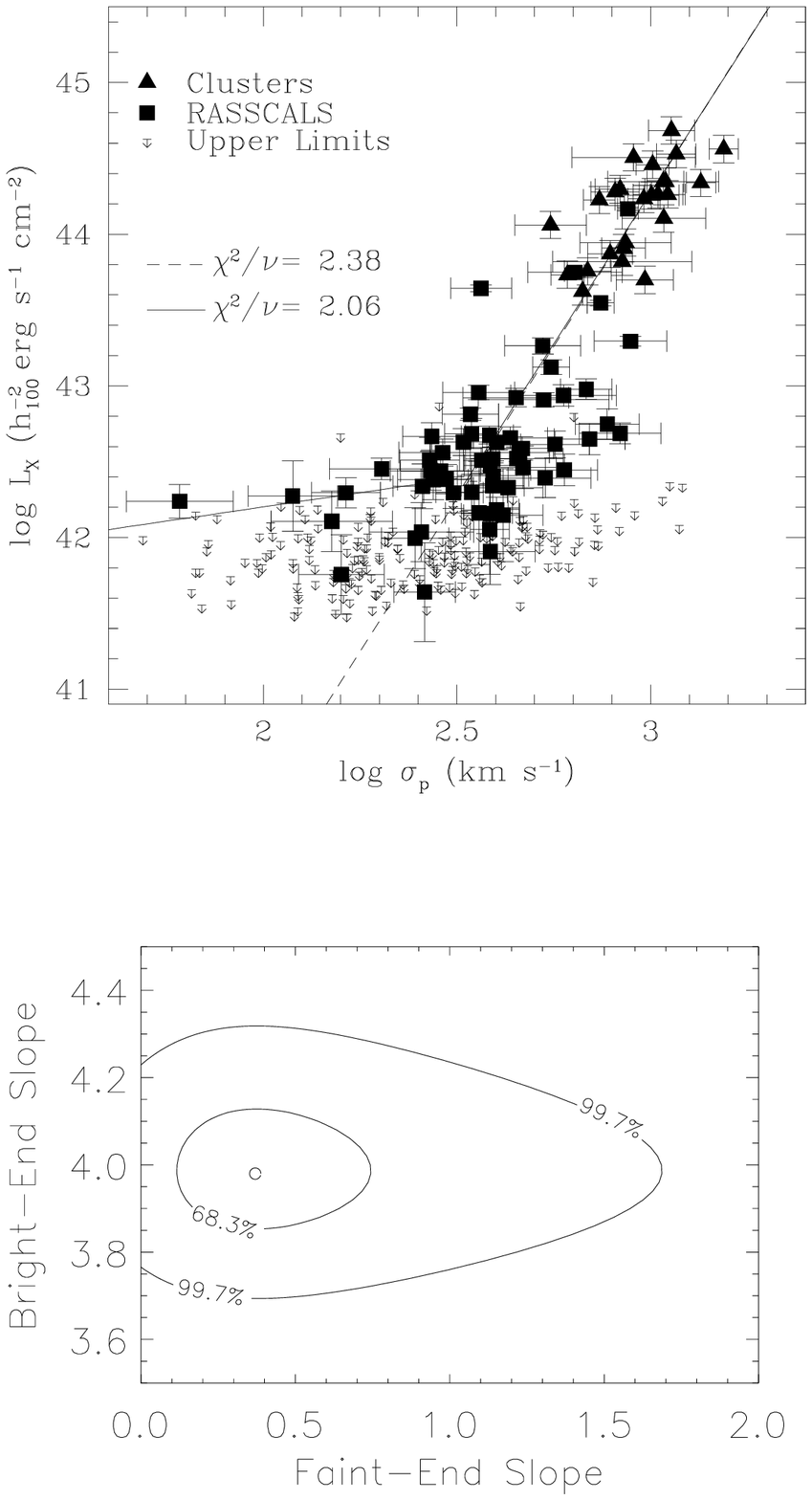}}
\end{center}
\figcaption[lxsig.eps]{The $L_X - \sigma_p$ relation. The dashed line
shows the single power law fit, and the solid line shows the broken
power law. The bottom panel shows the joint 68.3\% and 99.7\%
confidence intervals for the slopes of the broken power law. The
undetected RASSCALS are not considered in the fit, but in \S
\protect\ref{sec:broken} we consider the effect of including
the upper limits. \label{fig:lxsig}}
\end{figure*}

\begin{enumerate}
\item First, we fit a single power law by forcing $s_1 = s_2$ and
$\log{\sigma_k} = 0$, and applying the Press \etal (1995, \S 15.3)
package.  The result appears as the dashed line in Figure
\ref{fig:lxsig}.  We call this best-fit power law slope $s_0$.

\item Next, we vary the position of the knee of the power law over a
50 $\times$ 50 grid with bounds $\log{\sigma_k} = [2,3]$ and
$\log{L_k} = [42,44]$. At each point in the grid, we minimize the
$\chi^2$ over $s_1$ and $s_2$, using the Fletcher-Reeves-Polak-Ribiere
algorithm, which makes use of gradient information (Press \etal 1995,
\S 10.6). We start the minimization algorithm with $s_1 = s_2 = s_0$,
and require $0 < s_1 < 10$ and $0 < s_2 < 10$ as priors. This
procedure yields a function $\chi^2_\mr{min}(\sigma_k,L_k)$.

\item Finally, we minimize $\chi^2_\mr{min}(\sigma_k,L_k)$ to obtain
the position of the knee of the power law and the best-fit slopes
associated with it. 
The results of the fit appear in Figure
\ref{fig:lxsig}.

\end{enumerate}

We also try more robust estimators, such as the BCES bisector (Akritas
\& Bershady 1996). In general, these estimators are in good agreement
with the results of the $\chi^2$ fits; however, Akritas \& Bershady
(1996) do not provide a mechanism for assessing the quality of the
fit, and their package does not allow for the calculation of joint
two-dimensional confidence intervals. We therefore focus on the
$\chi^2$ statistic. Below we also consider how the fit changes with
the inclusion of the 199 upper limits.

\subsection{A Broken Power Law is the Best Fit}
\label{sec:broken}

Our data unambiguously favor a broken power law over a single power
law for the $L_X - \sigma_p$ relation. The confidence contours in
Figure \ref{fig:lxsig} show that the faint-end slope and the
bright-end slope are different at better than the 99.7\% confidence
interval. Furthermore, the scatter in the $L_X - \sigma_p$ relation is
actually reduced by fitting a broken power law instead of a single
power law.

The faint-end slope, $s_1 = 0.37 \pm 0.3$, is even shallower than the
earlier finding of Mahdavi \etal (1997), $s_1 = 1.56 \pm 0.25$, for
their fit to 9 low-luminosity RASSCALS. The shallowness of our
faint-end slope is remarkable because, unlike Mahdavi \etal (1997), we
remove sources of individual emission whenever possible, and model the
plasma temperature without fixing it at a particular value. The
bright-end slope, $s_2 = 4.02 \pm 0.1$, on the other hand, is
consistent with MZ98, whose $L_X - \sigma_p$ depends mainly on rich
clusters; their single power law fit has a slope $4.29 \pm 0.37$.  A
slight discrepancy is to be expected, because MZ98 use bolometric
X-ray luminosities, and we measure the luminosities in the 0.1--2.4
keV spectral range. However, this discrepancy should cause only an
$\Delta s_2 \approx 0.4$ offset in the slope for clusters with $L_X
\simgreat 10^{43}$ ergs s\m. Systems with $L_X \simless 10^{43}$ ergs s\m\
should have bolometric luminosities comparable to their 0.1--2.4 keV
luminosities.

We stress that our fitting procedure in no way favors the shallower
slope: we begin the $\chi^2$ minimization by setting both slopes equal
to the best-fit single power law. Also, a broken power law is the best
fit even if we exclude the lowest velocity dispersion group,
SRGb075, from the fit. Doing so, we would obtain $s_1 = 1.39 \pm 0.5$
and $s_2 = 3.99 \pm 0.3$.

Finally, we consider whether including the upper limits in the fit
changes the derived slopes. For this task we obtain the Astronomy
Survival Analysis Package (ASURV; Lavalley, Isobe, \& Feigelson 1992)
from http://www.astro.psu.edu/statcodes. ASURV implements the methods
described in Isobe, Feigelson, \& Nelson (1986) for regression of data
which includes both detections and upper limits. Furthermore, ASURV
allows for an intrinsic scatter in the relation.

For the 155 objects with $\sigma_p < 340$ km~s\m, we find that the
best-fit slope is $1.38 \pm 0.4$; for the 128 objects with $\sigma_p >
340$ km~s\m, it is $5.37 \pm 0.5$. Thus the inclusion of upper limits
does not bring the two slopes closer to each other; if anything, it
strengthens our claim that the $L_X - \sigma_p$ relation is best
described by a broken power law.

\section{Detection Statistics}

We now examine the statistical properties of the catalog in greater
detail. We seek a deeper understanding of the X-ray selection function
of the RASSCALS. A useful tool for this purpose is the number
distribution of a set of measurements $x$, which we label $N(x)$. For
example, we call the number distribution of the group velocity
dispersion $N(\sigma_p)$; the number distribution of the
distance-corrected group membership is N($n_\mr{17}$).

Although the traditional estimator of the number distribution is the
histogram, we compute $N(x)$ using the DEDICA algorithm (Pisani
1993). DEDICA makes use of Gaussian kernels to arrive at a
maximum-likelihood estimate of the number distribution. The resulting
smooth function, $N(x)$ is more useful than a histogram, because
$N(x)$ is nonparametric, and any structure within $N(x)$ is
statistically significant. We normalize $N(x)$ so that the number of
groups with $x_1 \le x \le x_2$ is given by,
\begin{equation}
\int_{x_1}^{x_2} N(x) dx.
\end{equation}

\subsection{An Abundance of Low $\sigma_p$ Systems}

Figure \ref{fig:obstats}a shows $N(\sigma_p)$ and $N(n_{17})$
separately for all RASSCALS and for those with significant X-ray
emission. Interestingly, $N(\sigma_p)$ for all \ntot\ groups is
double peaked; there are 102 RASSCALS (39\%) with $\sigma_p < 250$ km
s\m.

The abundance of these low $\sigma_p$ systems is puzzling considering
that many of them probably contain unrelated galaxies
(``interlopers'') (Frederic 1995), and that these interlopers
typically lead to an overestimate, not an underestimate, of the
velocity dispersion. Several plausible explanations for their frequent
occurrence exist.

\begin{enumerate}

\item It may be that the groups with low $\sigma_p$ are unbound,
chance projections along the line of sight. However, this situation is
highly unlikely for systems drawn from a complete redshift survey.
Chance superpositions in such a survey in fact have a larger mean
velocity dispersion than the bound groups do (Ramella \etal 1997).

\item The groups might be pieces of sheets or filaments in the large
scale distribution of matter. A collapsing sheet of galaxies which is
removed from the Hubble flow, and which is perpendicular to the line
of sight, might look like a group to the friends-of-friends algorithm.

\item The velocity dispersion $\sigma_p$ might not be related to the
mass distribution in a straightforward manner. For example, Mahdavi
\etal (1999) find that galaxy orbits in a subsample of the RASSCALS
have a significant mean radial anisotropy; and Diaferio (2000) uses
N-body simulations to show that systems with $\sigma_p \simless 300$
km~s\m\ have galaxy velocity dispersions that are uncorrelated with
the total group mass.
\end{enumerate}

\subsection{Detection Efficiency}
\label{sec:sim}

Here we use the groups we have detected as a basis for estimating (1)
the number of RASSCALS with X-ray emission too faint to be
observable by \ROSAT, and (2) the number of RASSCALS with no X-ray
emission, some of which might be unbound superpositions.

We begin by assuming that all the RASSCALS emit X-rays according to an
empirical relationship between $L_X$ and $\sigma_p$. We compute the
number of groups we expect to detect as a function of $\sigma_p$, and
compare the theoretical detection probability with the true detection
efficiency.

Suppose that all the RASSCALS emit X-rays according to a power law
relationship between $L_X$ and $\sigma_p$,
\begin{equation}
\log{L_X} = s \log{\sigma_p} + b.
\end{equation}
If the local flux detection threshold for a group at redshift
$z$ is $F_0$, it will be detectable if $\sigma_p > \sigma_0$,
where
\begin{equation}
\log{\sigma_0}  =  \frac{\log{\left[4 \pi F_0 (1 + z)^2 c^2 z^2 \right]} - b}{s}
\end{equation}
The theoretical probability that the group will be detected is then
\begin{equation}
P_\mr{th}(\sigma_p) = \int_{\log{\sigma_0}}^\infty p(\log{\sigma_p}) d \log{\sigma_p},
\end{equation}
where $p(\log{\sigma_p})$ is the probability distribution function of
$\log{\sigma_p}$. We calculate $P_\mr{th}(\sigma_p)$ for all 260
RASSCALS, taking $s = 4.02$ and $b = 32.19$ from the single power law
determined in \S \ref{sec:lxsig}. We approximate $\log{\sigma_p}$ as a
Gaussian with a standard deviation equal to 1.3 times the uncertainty
given in Table \ref{tbl:rasscals} for each group.  Multiplying the
error in $\sigma_p$ by 1.3 is a way of spreading the uncertainty in
the $L_X - \sigma_p$ relation directly into $P_\mr{th}(\sigma_p)$. The
resulting average theoretical probability of detecting a group with
velocity dispersion $\sigma_p$ is well approximated by
\begin{equation}
P_\mr{th}(\sigma_p) = \frac{1}{2} +
\frac{1}{2}\ {\mathrm{erf}} \left[ 4 \left(\log \frac{\sigma_p}{250\
{\mathrm{km\ s^{-1}}}} \right) \right],
\label{eq:erf}
\end{equation}
where ${\mathrm{erf}}(x)$ is the error function. 

\begin{figure*}
\resizebox{7in}{!}{\includegraphics{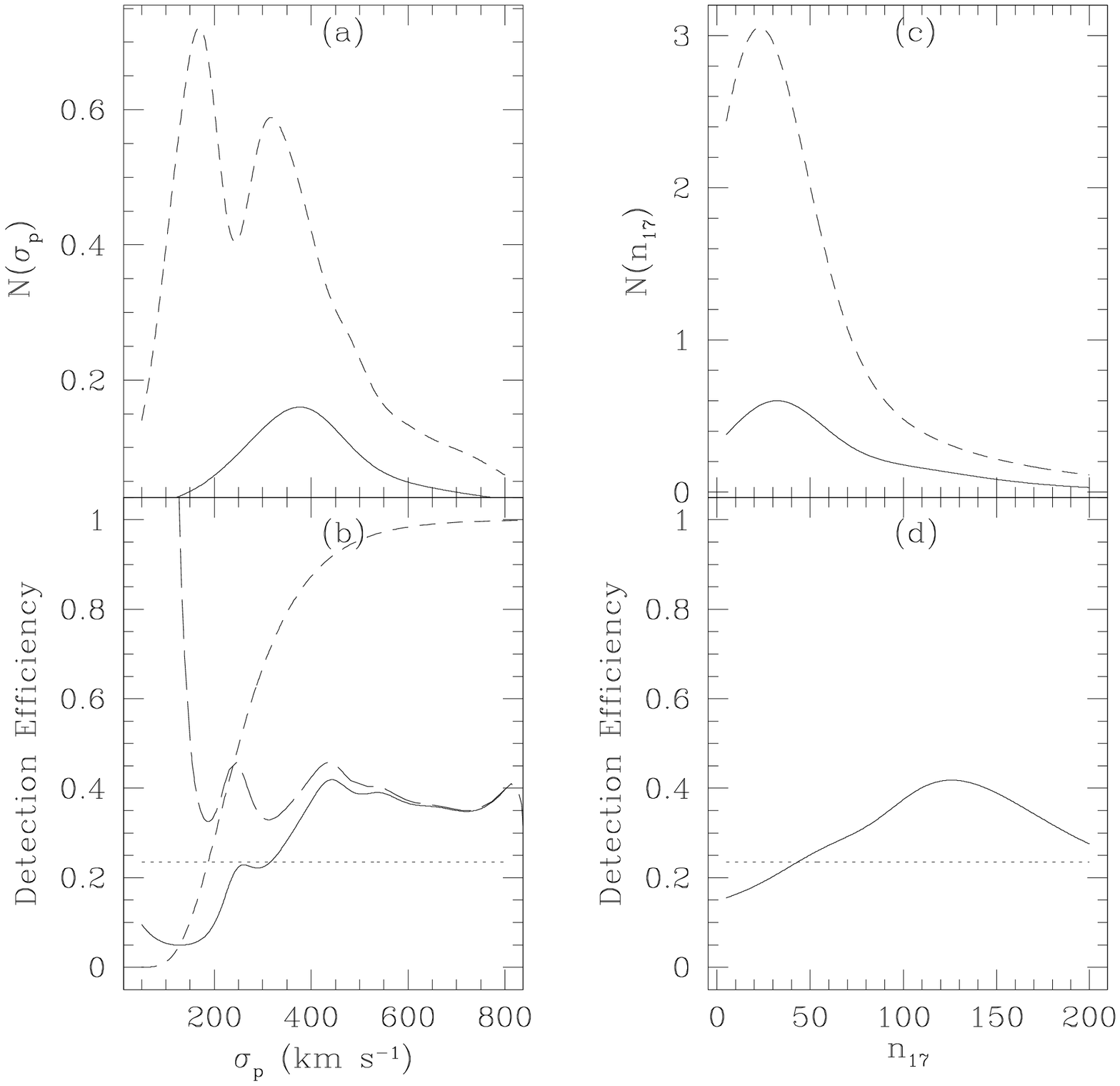}}
\figcaption[obstats.eps]{Number distributions of $\sigma_p$ and
$n_{17}$: (a) distribution of $\sigma_p$ for all groups (dashed line)
and for the $Q=A,B$ sample (solid line); (b) the observed detection
efficiency as a function of $\sigma_p$ (solid line), the expected
detection probability assuming $L_X \propto \sigma^4$ (short dashed
line), and their quotient (long dashed line). Similar analysis for
$n_{17}$, the number of system members brighter than an absolute
magnitude $M_z = -17$, appears in (c) and (d). The dotted horizontal
line shows the mean detection efficiency for the RASSCALS (23\%).
\label{fig:obstats}}
\end{figure*}

Figure \ref{fig:obstats}b shows the observed and the theoretical
detection probabilities. The solid line represents the fraction of the
RASSCALS we actually detect, $P_\mr{obs}(\sigma_p)$, and the short
dashed line shows $P_\mr{th}(\sigma_p)$, the fraction of the RASSCALS
we should detect given $L_X \propto \sigma^4$. The quotient,
\begin{equation}
f_X(\sigma_p) \equiv \frac{P_\mr{obs}(\sigma_p)}{P_\mr{th}(\sigma_p)},
\end{equation}
appears as the long dashed line. $f_X(\sigma_p)$ represents the
fraction of groups that should have extended X-ray emission in order
that we detect our set of 61 RASSCALS. Remarkably, $f_X$ is a nearly
constant $40\%$ for $\sigma_p > 150$ km s\m, and rises steeply for
$\sigma_p < 150$ km~s\m. The scatter around the theoretical
probability $P_\mr{th}(\sigma_p)$ introduces a $30\%$ uncertainty in
the breaking point $\sigma_p = 150$ km~s\m, but does not affect the
result that $f_X \approx 40\% \pm 8\%$ above the breaking point.

Thus Figure \ref{fig:obstats}b shows that we detect many fewer systems
overall than expected from the raw $L_X \propto \sigma_p^4$
relation. To match our observed detection efficiency, only $40\%$ of
groups with $\sigma_p > 150$ km~s\m\ must have extended X-ray
emission. The probability $f_X$ that a group contains X-ray emitting
gas does not seem to increase with the group velocity dispersion.

On the other hand, the detection of the $\sigma_p < 150$ km~s\m\
groups exceeds the expectation from $L_X \propto \sigma_p^4$. The
theoretical probability of detecting any of these low-$\sigma_p$
groups is near 0, and yet we detect 6\% of them. A flattening of the
true $L_X - \sigma_p$ relation for low velocity dispersion systems, of
the kind we discuss in \S \ref{sec:lxsig},  resolves the
discrepancy.

The result that only 40\% of the $\sigma_p > 150$ km~s\m\ RASSCALS
should emit X-rays has an interesting interpretation when combined
with the predictions of N-body (Frederic 1995, Diaferio 1999) and
geometric (Ramella \etal 1997) simulations of the local large-scale
structure. These simulations suggest that $\simgreat 80\%$ of groups
with $n \ge 5$ members drawn from a complete redshift survey should be
real, bound systems. If indeed $\simgreat 80\%$ of the RASSCALS are
bound, and our simulations are correct, then at least half the
bound groups must possess a negligible amount of extended X-ray
emission.

The X-ray data impose a lower limit of 40\%, and the simulations
impose an upper limit of 80\%, on the fraction of RASSCALS that are
real, bound systems of galaxies.

\section{Discussion}

The flattening of the $L_X - \sigma_p$ relation for systems with
$\sigma_p \simless 340$ km~s\m\ is now well established, not just by
our study, but by pointed \ROSAT observations of an independent sample
of 24 groups (Helson \& Ponman 2000). This similarity breaking is
particularly striking, because it is in conflict with the $L_X - T$
relation for systems of galaxies, which actually steepens as the
temperature drops (Metzler \& Evrard 1994; Ponman \etal 1996).

It is difficult to dismiss the $L_X - \sigma_p$ flattening by claiming
that the velocity dispersion of the discrepant groups is biased
towards lower values. Most groups drawn from redshift surveys contain
unrelated galaxies which tend to bias $\sigma_p$ towards larger values
(Frederic 1995; Diaferio 1999).  It is also no longer possible to
argue (e.g. Mulchaey \& Zabludoff 1998) that the flattening is due to
a failure to remove detectable contamination.  We, as well as Helson
\& Ponman (2000), remove such contamination to the extent allowed by
the data.

The shallow $L_X - \sigma_p$ slope may be explainable through the
``mixed-emission'' scenario proposed by Dell'Antonio \etal (1994). It
is possible that a number of galaxies with faint, X-ray emitting ISMs
are embedded within the intragroup medium. These individually emitting
galaxies could contribute significantly to the total luminosity and
place it above the virial value. A large fraction of such emission
would be neither directly detectable nor removable, appearing instead
as fluctuations in excess of Poisson and instrumental noise superposed
on the central emission peak (Soltan \& Fabricant 1990). Further
verification of the mixed emission scenario thus depends on higher
quality observations of the lowest velocity dispersion systems with
the \emph{Chandra} or \emph{XMM} missions.

However, we can investigate whether the break in the $L_X - \sigma_p$
relation is linked to other physical properties of the RASSCALS.  One
possibility is that the excess emission is characteristic of the
dynamically youngest groups, those perhaps still in the process of
formation. An indicator of such a dynamical state might be the
fraction of spiral member galaxies, $f_\mr{sp}$. If the dominant
process for the formation of elliptical galaxies in groups is
galaxy-galaxy interaction, then one might expect a system with
$f_\mr{sp} \approx 0$ to be much more evolved than a group composed
mainly of spiral galaxies.

\begin{figure*}
\resizebox{7in}{!}{\includegraphics{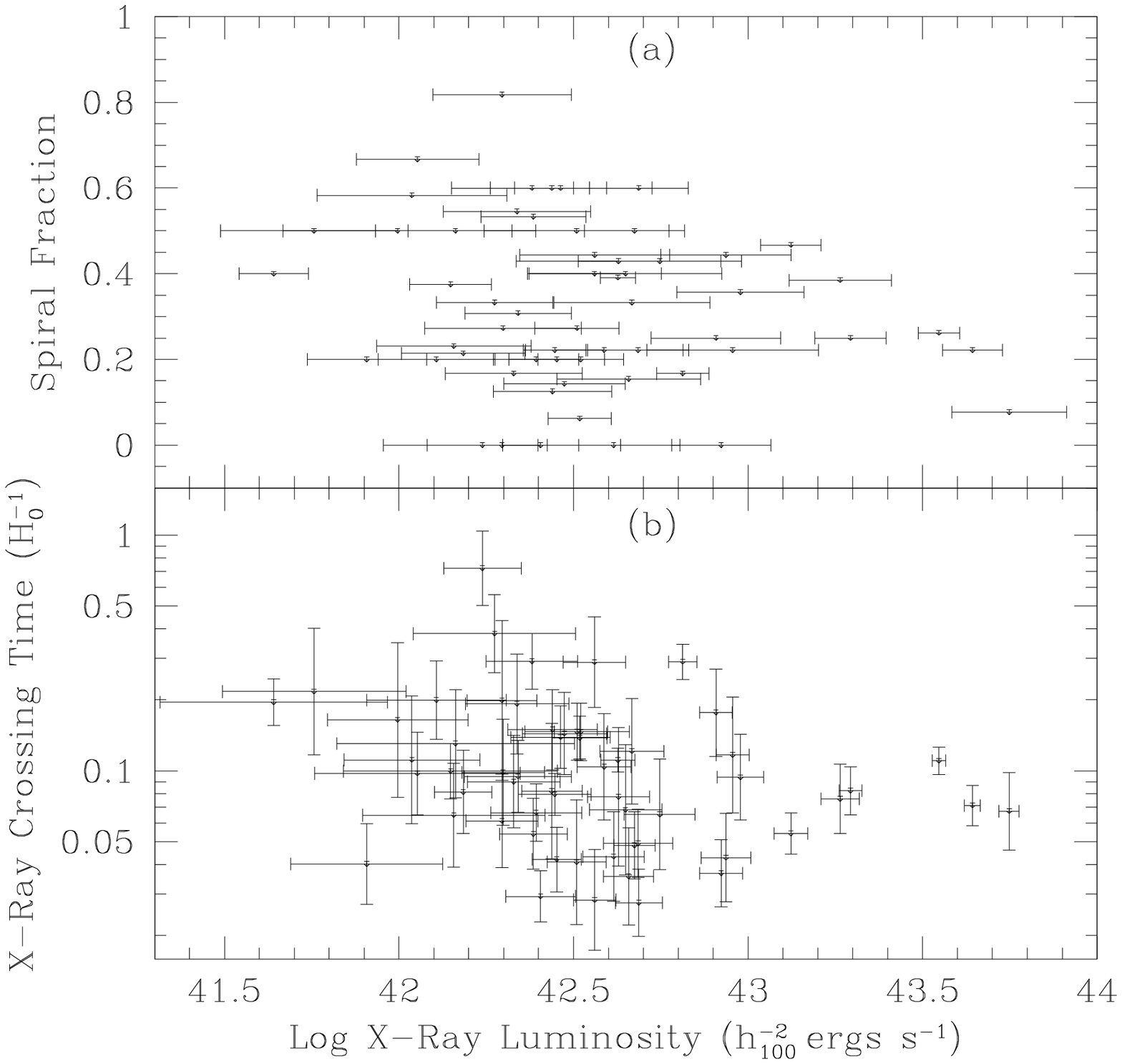}}
\figcaption{Correlation of the spiral fraction $f_\mr{sp}$ (a), and
the X-ray crossing time $R_\xi H_0 / \sigma_p$ (b), with the 0.1--2.4
keV X-ray luminosity.
\label{fig:corrlum}}
\end{figure*}

Figure \ref{fig:corrlum}a shows a weak correlation between the X-ray
luminosity and the spiral fraction. The correlation between the two
quantities is barely significant (Kendall's $\tau = -0.109$, with $P =
0.22$, a 1-$\sigma$ result). It is noteworthy that no system with
$f_\mr{spi} \ge 0.5$ is more luminous than $5\hhh^{-2} \times 10^{42}$
ergs s\m---groups that are spiral-dominated tend to have below average
X-ray luminosities.

However, closer inspection reveals that the spiral fraction is not
related to the $L_X - \sigma_p$ flattening. SRGb075, the X-ray
emitting group with the smallest velocity dispersion ($\sigma_p = 60$
km~s\m) has $f_\mr{spi} = 0.2$. The group with the next smallest
$\sigma_p$, SS2b293, has $f_\mr{sp} = 0.33$, and the following group,
NRGb045, has $f_\mr{sp} = 0.2$. Although a trend relating $f_\mr{sp}$
and $L_X$ probably exists, the RASSCALS that are responsible for the
flattening of the $L_X - \sigma_p$ relation have spiral fractions
comparable to those of higher velocity dispersion groups.

Another possible indicator of the dynamical age of a system of
galaxies is its crossing time. Groups where galaxies have completed
many orbits might be closer to dynamical equilibrium than those where the
galaxies have made only a few crossings. We note, however, that if the
accretion of external galaxies plays a significant role in the
evolution of a group, it may not reach dynamical equilibrium even
after many crossing times have passed (Diaferio \etal 1993).

The crossing time of a system in units of the Hubble time is roughly
$t_c = R H_0 / \sigma_p$, where $R$ is the characteristic
size. Because many of the groups in our sample have fewer than 9
members, computing $R$ from the optical data is likely to lead to
large errors in $t_c$.  Instead, we use the NOCORE radius (\S
\ref{sec:nocore}) to estimate the crossing time, $t_c = R_\xi H_0 /
\sigma_p$. The NOCORE radius provides a characteristic scale for the
X-ray emission, and hence for the gravitational potential of each
group.

There is a significant correlation between $L_X$ and $t_c$ in Figure
\ref{fig:corrlum}b (Kendall's $\tau = -0.228$, $P = 0.01$, a nearly
3-$\sigma$ result). Of course, this effect follows directly from the
relationship between $L_X$ and $\sigma_p$ (which exhibit a 10-$\sigma$
correlation). Because $R_\xi$ is uncorrelated with $\sigma_p$, the
inclusion of $R_\xi$ increases the scatter.

However, the $L_X - t_c$ comparison does reveal an interesting
property of the groups which contribute to the flattening of the $L_X
- \sigma_p$ correlation. These groups have $t_c > 0.3 H_0^{-1}$; they
have longer crossing times than the groups in the steeper, $L_X
\propto \sigma^4$ portion of the relation.  Thus we have an indication
that the X-ray overluminous groups are also the ones where the
crossing time is a large fraction of the Hubble time. An explanation
of this result in terms of the dynamical histories of the
low-$\sigma_p$ groups awaits a much deeper optical and X-ray probe of
their structure.

\section{Conclusion}

The RASSCALS are the largest extant combined X-ray and optical catalog
of galaxy groups. We draw the systems from two redshift surveys that
have a limiting magnitude of $m_z = 15.5$ and cover $\pi$ ster of the
sky. There are \ntot\ systems, of which 23\% have statistically
significant X-ray emission in the \ROSAT\ All-Sky Survey after we
remove contamination from unrelated sources. We include a catalog of
the systems.

We calculate the X-ray selection function for our sample. The behavior
of the function implies that only 40\% of the RASSCALS are
intrinsically X-ray luminous. The remaining $\approx $ 60\% of the
RASSCALS are either chance superpositions, or bound systems devoid of
hot gas.

We examine the relationship between the X-ray luminosity $L_X$ and the
velocity dispersion $\sigma_p$ for the 59 high-quality RASSCALS and a
representative sample 25 of rich clusters not internal to our
data. The best fit relation is a broken power law with $L_X \propto
\sigma_p^{0.37 \pm 0.3}$ for $\sigma_p < 340$ km~s\m, and $L_X \propto
\sigma_p^{3.9 \pm 0.1}$ for $\sigma_p > 340$ km~s\m. Whether we
include the upper limits in our analysis, or assume a dominant
intrinsic scatter in the relation, a broken power law with a shallow
faint-end slope is still a better fit than a single power law. 

Stressing that we have been careful to remove contamination from
individual galaxies and unrelated sources, we conclude that the
flattening in the $L_X - \sigma_p$ relation for groups of galaxies is
a physical effect. A potential mechanism for the excess luminosity of
the faintest systems is the ``mixed emission'' scenario (Dell'Antonio
\etal 1994): the emission from the intragroup plasma may be
irrecoverably contaminated by a superposition of diffuse X-ray sources
corresponding to the hot interstellar medium of the member galaxies. A
final explanation of the flattening of the $L_X - \sigma$ relation
must focus on the detailed X-ray and optical structure of the
groups with small velocity dispersions ($\sigma_p < 150$ km~s\m).

We plan to calculate the X-ray luminosity function of the RASSCALS
soon. Deep optical spectroscopy of these systems is already underway,
and the first results appear in Mahdavi \etal (1999).

We are grateful to the anonymous referee and the editor, Gregory
Bothun, for comments which led to significant improvement of the
paper. We thank Saurabh Jha and Trevor Ponman for useful discussions.
This research was supported by the National Science Foundation
(A. M.), the Smithsonian Institution (A. M., M. J. G.), and the
Italian Space Agency (M. R.).

\appendix
\section{Luminosity, Flux and Temperature Calibration}
\label{sec:calib}

Here we describe the procedure we use to convert the \ROSAT\ PSPC
count rate into a 0.1--2.4 keV X-ray luminosity. Our procedure does
not require fixing or guessing the plasma temperature.  Instead, we
fold the uncertainty in the temperature directly into the derived
luminosities.

The decontaminated source count rate within a ring of projected radius
$R$ is
\begin{equation}
S = \frac{\pi R^2}{\pi R^2 - A_{\mr{clean}}} \left( \sum_i
\frac{N_i}{E_i}\right) - \pi R^2 B,
\end{equation}
where $A_\mr{clean}$ is the area of the portion of the ring removed
during the decontamination process, $N_i$ is the total count rate
within pixel $i$, $E_i$ is the exposure time within pixel $i$, $B$ is
the average background count rate per unit area on the sky, and the
summation is over all pixels within the ring. Pixels which fall
partially inside the ring are appropriately subdivided. The error
in the source count rate, $\sigma_S$, is given by
\begin{equation}
\sigma_S^2 = \left( \frac{\pi R^2}{\pi R^2 - A_{\mr{clean}}} \right)^2
\left( \sum_i \frac{N_i}{E_i^2} \right) + \left (\pi R^2 \sigma_B \right)^2, 
\end{equation}
where $\sigma_B$ is the uncertainty in the background.

Because all our systems have redshift $z < 0.04$, the 0.1--2.4 keV
X-ray luminosity, $L_X$, is
\begin{equation}
\label{eq:lx}
L_X = 4 \pi F (1 + z)^2 \left( \frac{c z}{H_0} \right)^2.
\end{equation}
The 0.1--2.4 keV flux, $F$ from the GPT is then
\begin{equation}
\label{eq:flux}
F = C(N_H,T) S.
\end{equation} 
Here $C(N_H,T)$ is a function suited to the \ROSAT\ PSPC Survey Mode
instrumental setup which converts the 0.5--2.0 keV count rate to the
appropriate 0.1--2.4 keV flux from a Raymond \& Smith (1977) spectrum
with the abundance fixed at $30$\% of the solar value. $C(N_H,T)$
depends on $N_H$, the total hydrogen column density along the line of
sight, which we compute using the results of Dickey \& Lockman (1990),
and the emission-weighted plasma temperature, $T$. 

We cannot accurately determine $T$ independently of $F$ from the RASS
data. However, once $N_H$ is fixed, $C(N_H,T)$ varies only 15\%--20\%
for $ 0.3 $ keV $\le kT \le 10$ keV. We therefore fold this
uncertainty in $T$ into our calculation of the flux.

If $p_C(C)$ is the probability distribution function (PDF) of
$C(N_H,T)$, and $p_S(S)$ is the PDF of the source count rate $S$,
then the PDF of the flux is (Lupton 1993, pp. 9--10)
\begin{equation}
p_F(F) = \int_0^\infty p_C(C) p_S(F/C) \frac{dC}{C}
\end{equation}
If the PDF of the group's emission-weighted temperature is $p_T(T)$,
then, by the law of transformation of probabilities,
\begin{equation}
p_C(C) = p_T(T) \left| \frac{dT}{dC} \right|.
\end{equation}
Approximating $p_S(S)$ as a Gaussian distribution with mean $S$ and
standard deviation $\sigma_S(S)$, we obtain
\begin{equation}
p_F(F) = \frac{1}{\sqrt{2 \pi} \sigma_S}
\int_0^\infty \frac{p_T(T)}{C(N_H,T)} \exp \left[ -\frac{1}{2} 
\left( \frac{F/C(N_H,T) - S}{\sigma_S} \right)^2 \right] dT.
\end{equation}
We take $p_T(T)$ to be $(9.7\ \mr{keV})^{-1}$ over the range 0.3--10
keV, and zero everywhere else. We have also tried a more sophisticated
approach, with $p(T)$ proportional to the observed temperature
function of systems of galaxies (Markevitch 1998). The difference
between the resulting PDF and the constant $p_T(T)$ PDF is negligible
compared with the error introduced by the uncertainty in the
temperature function itself.

\renewcommand{\thetable}{\arabic{table}}

\begin{deluxetable}{lr@{ --- }rr@{ --- }rc}
\tablecaption{Sky Coverage of the Optical Group Catalog}
\tablewidth{0in}
\tablehead{\colhead{Field Name} & \multicolumn{2}{c}{ $\alpha_{2000}$ } &
\multicolumn{2}{c}{$\delta_{2000}$} }
\startdata
NRG  & 8.5 hr & 17 hr & 8.5\degr &43.5\degr  \\
SRG  & 21.5 hr & 3 hr & -2\degr  &32\degr  \\
SS2\tablenotemark{a} & 21 hr & 5 hr & -40\degr  & 1.5\degr  \\
SS2\tablenotemark{b} & 10 hr & 15 hr & -26\degr & 0\degr  \\
\enddata
\tablenotetext{a}{First portion, excluding sections with galactic
latitude $b < 40$\degr.}
\tablenotetext{a}{Second portion, excluding sections with galactic
latitude $b < 40$\degr.}
\label{tbl:skyportions}
\end{deluxetable}

\newcommand{\mcfour}{\multicolumn{2}{c}{\tablenotemark{d}}}
\newcommand{\mcone}{\multicolumn{1}{c}{\nodata}}

\begin{deluxetable}{crrrrr@{$\pm$}lr@{$\pm$}lr@{$\pm$}ll}
\tablecaption{The RASSCALS: Basic Properties}
\tablewidth{0in}
\tablefontsize{\small}
\tablehead{ \colhead{RASSCALS} & \colhead{$\alpha$\tablenotemark{a}} 
& \colhead{$\delta$\tablenotemark{a}} & \colhead{$n$} & \colhead{$n_{17}$} &
 \multicolumn{2}{c}{$c z$}
& \mc{$\log{\sigma_p}$}& \mc{$\log{L_X}$\tablenotemark{b}}
& \colhead{Other} \nl 
\colhead{ID} & \colhead{J2000} & \colhead{J2000} & & &
 \multicolumn{2}{c}{km s\m} & \mc{km s\m} & \mc{$\hhh^{-2}$ erg s\m} &
 \colhead{Identification\tablenotemark{c}}}
\startdata
SS2b003   &00:08:55.8 &$-$37:28:15&    5&  35& 8357&  60& 2.14 & 0.20&  \mc{$<42.1$} & \nodata \nl
SRGb061   &00:11:44.7 &$+$28:21:35&   10&  57& 7855& 163& 2.71 & 0.06&  \mc{$<41.9$} & PPS058     \nl
SS2b004   &00:14:47.2 &$-$07:14:16&    5&  11& 5290&  65& 2.19 & 0.14&  \mc{$<41.7$} & \nodata    \nl
SS2b005   &00:15:31.5 &$-$24:07:38&    5&  24& 7390&  35& 1.86 & 0.15&  \mc{$<42.0$} & \nodata    \nl
SRGb062   &00:18:25.2 &$+$30:04:13&   13&  49& 6811& 122& 2.64 & 0.10& 42.66 & 0.07   & MGBR       \nl
SRGb063   &00:21:38.4 &$+$22:24:20&   10&  25& 5665&  87& 2.46 & 0.11& 42.56 & 0.09   & PPS062     \nl
\nodata   & \mcone  &  \mcone & \mcone  &\mcone&\mc{\nodata}&\mc{\nodata}
&\mc{\nodata} & \mcone\\
\nodata   & \mcone  &  \mcone & \mcone  &\mcone&\mc{\nodata}&\mc{\nodata}
&\mc{\nodata} & \mcone\\
\nodata   & \mcone  &  \mcone & \mcone  &\mcone&\mc{\nodata}&\mc{\nodata}
&\mc{\nodata} & \mcone\\
\enddata
\tablenotetext{a}{For the 59 groups with a listed X-ray luminosity,
we report the X-ray centroid; for the others we report the mean RA and DEC
of the galaxies in the group.}
\tablenotetext{b}{Luminosity in the 0.1--2.4 keV band within an aperture of 0.5$\hhh^{-1}$ Mpc.}
\tablenotetext{c}{By no means complete. A: Abell cluster; AWM: Albert \etal (1997)
Groups; HCG: Hickson Compact Groups (Hickson 1982); MGBR: Studied in
greater detail in Mahdavi \etal (1999); PPS: Loose Groups in the
Perseus-Pisces Survey (Trasarti-Battistoni 1998); ZM: Zabludoff \&
Mulchaey (1998) Poor Groups}
\tablecomments{The complete table will be available in
the electronic version of \emph{The Astrophysical Journal}.  The first
few lines are shown to elucidate form and content.}
\label{tbl:rasscals}
\end{deluxetable}

\begin{deluxetable}{cr@{$\pm$}lr@{$\pm$}lr@{$\pm$}lr@{$\pm$}lr@{$\pm$}lr@{$\pm$}ll}
\tablecaption{The RASSCALS: Detailed Properties}
\tablewidth{0in}
\tablefontsize{\small}
\tablehead{ \colhead{RASSCALS} & \mc{$\log{\sigma_p}$} & 
\mc{$\log{L_X(0.25)}$} & 
\mc{$\log{L_X(0.5)}$} & \mc{$\log{L_X(R_\xi)}$} &
\mc{$R_\xi$} & \mc{$\log{t_c}$} & \colhead{$f_\mr{sp}$}   \nl
\colhead{ID} &  \mc{km s\m} &
\multicolumn{6}{c}{$\hhh^{-2}$ erg s\m} &
\mc{$\hhh^{-1}$ Mpc} & \mc{$H_0^{-1}$}  & \colhead{} }
\startdata
SRGb062 & 2.64 & 0.10 & 42.41 &  0.08 & 42.66 &  0.07 & 42.27 &  0.08 & 0.15 & 0.07 & -1.45 & 0.21 & 0.15 \\
SRGb063 & 2.46 & 0.11 & 42.32 &  0.09 & 42.56 &  0.09 & 42.71 &  0.10
& 0.84 & 0.30 & -0.54 & 0.19 & 0.40 \\
\nodata & \mc{\nodata} &\mc{\nodata} &\mc{\nodata} &\mc{\nodata} &
\mc{\nodata} &\mc{\nodata} & \mcone  \\
\nodata & \mc{\nodata} &\mc{\nodata} &\mc{\nodata} &\mc{\nodata} &
\mc{\nodata} &\mc{\nodata} & \mcone  \\
\nodata & \mc{\nodata} &\mc{\nodata} &\mc{\nodata} &\mc{\nodata} &
\mc{\nodata} &\mc{\nodata} & \mcone  \\
\enddata
\tablecomments{Data for the detected groups, excluding the optically
spoiled clusters NRGs372 and NRGs392. 
The 0.1--2.4 keV X-ray luminosities are computed within
$0.25\hhh^{-1}$ Mpc and $0.5\hhh^{-1}$ Mpc, as well as within the
NOCORE radius, $R_\xi$. The crossing time $t_c = R_\xi H_0 /
\sigma_p$, and $f_\mr{sp}$ is the fraction of group members that are
spiral galaxies.}
\tablecomments{The complete table will be available in
the electronic version of \emph{The Astrophysical Journal}.  The first
few lines are shown to elucidate form and content.}
\label{tbl:detailed}
\end{deluxetable}

\begin{deluxetable}{rrcr@{$\pm$}l}
\tablecaption{Clusters Data}
\tablewidth{0in}
\tablehead{\colhead{Name} & \colhead{z} & 
\colhead{$\log{L_X}$\tablenotemark{a}} & 
\mc{$\log{\sigma_p}$\tablenotemark{b}} \nl 
 & & \colhead{$\hhh^{-2}$ erg s\m} & \mc{km s\m}}
\tablefontsize{\normalsize}
\startdata
     A85 & 0.052 & 44.28 & 2.99 & 0.02\nl
    A119 & 0.044 & 43.94 & 2.83 & 0.06\nl
    A399 & 0.072 & 44.23 & 3.05 & 0.04\nl
    A401 & 0.074 & 44.46 & 3.06 & 0.03\nl
    A478 & 0.088 & 44.51 & 2.96 & 0.16\nl
    A754 & 0.054 & 44.35 & 2.82 & 0.04\nl
   A1651 & 0.085 & 44.26 & 3.00 & 0.07\nl
   A1736 & 0.046 & 43.70 & 2.98 & 0.07\nl
   A1795 & 0.062 & 44.30 & 2.92 & 0.04\nl
   A2029 & 0.077 & 44.53 & 3.07 & 0.03\nl
   A2065 & 0.072 & 44.10 & 3.03 & 0.11\nl
   A2142 & 0.089 & 44.68 & 3.05 & 0.04\nl
   A2256 & 0.058 & 44.34 & 3.13 & 0.02\nl
   A2319 & 0.056 & 44.56 & 3.19 & 0.02\nl
   A3112 & 0.070 & 44.06 & 2.74 & 0.06\nl
   A3266 & 0.055 & 44.26 & 3.04 & 0.03\nl
   A3376 & 0.046 & 43.76 & 2.84 & 0.04\nl
   A3391 & 0.054 & 43.87 & 2.82 & 0.10\nl
   A3395 & 0.050 & 43.91 & 2.93 & 0.03\nl
   A3558 & 0.048 & 44.23 & 2.99 & 0.02\nl
   A3571 & 0.040 & 44.26 & 3.02 & 0.04\nl
   A3667 & 0.053 & 44.35 & 2.99 & 0.02\nl
   A4059 & 0.048 & 43.82 & 2.93 & 0.18\nl
Cygnus A & 0.057 & 44.30 & 3.20 & 0.11\nl
   MKW3S & 0.045 & 43.73 & 2.79 & 0.04\nl 
\enddata
\tablenotetext{a}{The X-Ray luminosities are in the 0.1--2.4 keV band,
from Markevitch (1998). We take uncertainty in 
the luminosities to be 20\%.}
\tablenotetext{b}{From Fadda \etal (1996).}
\label{tbl:clusters}
\end{deluxetable}

\end{document}